\documentclass[fleqn,10pt]{wlscirep}
\usepackage[utf8]{inputenc}
\usepackage[T1]{fontenc}
\usepackage[english]{babel}
\usepackage{mathtools}
\usepackage{amsmath}
\usepackage{longtable}
\usepackage{amsmath}
\usepackage{eucal}

\title{Link Prediction via Graph Attention Network}

\author[1]{Weiwei Gu}
\author[1]{Fei Gao}
\author[1]{Xiaodan Lou}
\author[1,*]{Jiang Zhang}
\affil[1]{School of Systems Science, Beijing Normal University, Beijing 100875, P.R.China}
\affil[*]{zhangjiang@bnu.edu.cn}

\begin{document}

\begin{abstract}
Link prediction aims to infer missing links or predicting the future ones based on currently observed partial networks, it is a fundamental problem in network science with tremendous real-world applications. However, conventional link prediction approaches neither have high prediction accuracy nor being capable of revealing the hidden information behind links.
To address this problem, we generalize the latest techniques in deep learning on graphs and present a new link prediction model - DeepLinker. Instead of learning node representation with the node label information, DeepLinker uses the links as supervised information. Experiments on five graphs show that DeepLinker can not only achieve the state-of-the-art link prediction accuracy, but also acquire the efficient node representations and node centrality ranking as the byproducts. Although the representations are obtained without any supervised node label information, they still perform well on node ranking and node classification tasks.
\end{abstract}

\maketitle

\section{Introduction}
Many real world data come naturally in the form of pairwise relations, such as protein-protein interaction in human cell, paper citations in scientific research and drug-target interaction in medicine discovery \cite{chen2012drug,campillos2008drug,zhou2016ranking}. These linkages contain rich information on node properties, network structures and network evolution. To predict the existence of a relation is a fundamental task in network science and of great importance in practice. For networks in biology like protein-protein interaction network, metabolic network, and food webs, the discovery and validation of links require significant experimental effort. Instead of blindly checking all possible links, link prediction can help scientists to focus on the most likely links and thus sharply reduce the experimental cost. For WWW, social networks, and citation networks, link prediction can help in recommending relevant pages, finding new friends, or discovering new citations \cite{craven2000learning,popescul2003statistical,liben2007link}.

The conventional link prediction methods can be divided into several categories. Local link prediction approaches make prediction based on the assumption that two nodes are more likely to be connected if they have many common neighbors \cite{barabasi1999emergence,zhou2009predicting}. These local similarity based methods are fast and highly parallel since they only consider local network structure. But their prediction accuracy are very low especially when networks are sparse and large. The global link prediction approaches take the whole network's structural similarity into consideration ~\cite{liu2013hidden,zhou2009predicting,rucker2012network}, those methods have higher link prediction accuracy compared with the local ones, but they have high computational complexity problem which prevent them to be applied on graphs that contain million and billion nodes. There are also some probabilistic and statistical based approaches that assuming there is a known prior structure of the network, like hierarchical or circles structures \cite{clauset2008hierarchical,huang2006link}. But those methods can not get over the problem of low link prediction accuracy. Besides, we can hardly extract the hidden network structure and node properties from above mentioned conventional link prediction approaches.

Recently, there has been a surge of algorithms that seek to make link prediction through network representation learning which automatically decoding local and global structural information from graphs. The idea behind these algorithms is to learn a mapping function that embedding nodes as points in a low-dimensional space $\mathbb{R}^d$ with nodes vector represent the information extracted from the original graph. Most of the network representation based methods are based on the Skip-Gram method or matrix factorization, such as DeepWalk, node2vec, LINE, and struc2vec \cite{perozzi2014deepwalk, grover2016node2vec, ou2016asymmetric, tang2015line, ribeiro2017struc2vec}. Those algorithms are task agnostic, the learned representations then used to perform graph-based downstream machine learning tasks, such as node classification, node ranking as well as link prediction \cite{gu2017hidden,grover2016node2vec}. Compared with the conventional ones those representation based methods have achieved a much higher link prediction accuracy. But they still have several drawbacks, to start with, there is no supervised information during the training process, we can't evaluate the embedding quality unless performing the downstream machine learning tasks. Second, nodes' representation vectors are updated directly without considering the dynamic changing structure of the networks, network structures are not static, the nodes and edges are changing rapidly. Those algorithms can't assign meaningful vector to a newly added node. Moreover, the expressive power is limited, because the embedding process is fixed by the random walk strategy\cite{perozzi2014deepwalk, grover2016node2vec,ribeiro2017struc2vec}. Finally, the representations can be hardly extent for inductive learning since the embedding vectors can not be transferred to similar graphs\cite{hamilton2017inductive}.

More recently, deep learning techniques based on neural networks have achieved triumphs in image processing \cite{he2016deep} and natural language processing \cite{gehring2016convolutional}. This stimulates the extensions of the methods on graph structures to perform node classification and link prediction tasks by converting the network structures into a low dimensional representations. For example, graph convolutional network (GCN) \cite{kipf2016semi} borrows the concept of convolution from the convolutional neural network (CNN) and convolve the graph directly according to the connectivity structure of the graph. After that, Velickovic et al.\cite{velickovic2017graph} further proposed graph attention networks (GAT) and obtained the state-of-art accuracy in node classification task. Following the self-attention mechanism, GAT compute representation of each node by combining its neighborhoods vectors in an adaptive way. The attention here is an adjustable weights on different neighbor nodes which can be updated dynamically according to the states of the nodes within a local connected neighborhood.

Nevertheless, those algorithms mentioned above and their extensions \cite{NIPS2016_6081} have the scalability problem since they take the whole graph as input and recursively expand neighborhoods across layers. This expansion is computationally expensive especially when graphs grow large. Due to the scale free property in many graphs, when hub nodes are sampled as the 1st-order neighbors, their 2nd-order neighbors usually quickly fill up the memory that usually leads to the memory bottleneck problem. This problem prevents GAT and GCN to be applied on large scale networks. 

GraphSAGE \cite{hamilton2017inductive} tries to solve the memory bottleneck problem by sampling a fixed-size neighborhood during each iteration, after that it performs a specific aggregator over feature extractor. The sampling strategy in GraphSAGE yields impressive performance on node labeling task over several large-scale networks. FastGCN \cite{chen2018fastgcn} then proposes to view GCN \cite{kipf2016semi} as integral transforms of embedding functions under probability measure. The classification accuracy is highly comparable with the original GCN while gets rid of the reliance on the test data.

However, most of the graph convolution based methods mainly apply node labels instead of linkages as the supervised information. Node labels, however, are always scarce in most real networks, there are only several types of networks that have node label information, such as the Protein-Protein interactive network, citation networks and so on. Besides, linkages rather than node attributes contain much richer information about network structure and evolution. According to the similarity and popularity theory \cite{papadopoulos2012popularity}, the links within a network not only contain nodes similarity \cite{csimcsek2008navigating} but also encode nodes popularity information\cite{barabasi1999emergence}. Take the formation of a citation network for example, papers tend to cite literatures that not only have higher content similarity but also have greater popularity  \cite{wu2014generalized}. Linkages contains much more structure information than node labels. And the representation power is influenced strongly by the supervised information, thus, for most downstream machine learning tasks such as link prediction, visualization and community detection, linkages rather than node labels should be used as the supervised information because they encodes at least both popularity and similarity.


In this paper, we propose DeepLinker, a new model that extends the GAT architecture to be applied on predicting the missing links over various networks. By adopting the attention mechanism, we can not only make predictions on links but also learn meaningful node representations that can be used to identify node categories especially when the labeled nodes are spares. The learned attentional weights can help us form directed and weighted graphs from the un-directed and unweighted ones. The attentional weights also show potential in evaluating nodes' importance and measuring node centrality. The original GAT model can not be directly applied on the supervised link prediction task due to the following reasons. First, the complexity of classification task is $\mathcal{O}$($N$), while the complexity of link prediction is $\mathcal{O}$($N^{2}$), where $N$ is the number of nodes. Compared with node classification, link prediction usually involves a much larger node feature computation, which bringing the memory bottleneck problem. Second,the original GAT model needs access to the entire network while performing node classification inference, but we can not directly apply this strategy to link prediction task due to the well known scale-free property in real networks since the expansion of the neighborhood of a hub node can quickly fill up a large portion of the graph. Last, although large mini-batches are preferable to reduce the communication cost, they may slow down the convergence rate in practice \cite{byrd2012sample} since the decrease in mini-batch size typically increases the rate of convergence \cite{li2014efficient} in the optimization process. Whereas, in the original GAT model, a small mini-batch usually involves a large amount of nodes, which decreases the convergence rate and usually leads to poor performance in link prediction accuracy.

Here we solve the memory bottleneck and the convergence problem by incorporating the mini-batch strategy via sampling a fixed neighborhoods\cite{hamilton2017inductive}. The difference between DeepLinker and GraphSAGE lies in the sampling stratagy. In DeepLinker we sample only once and fix the sampling neighborhoods during the training process, GraphSAGE keeps changing neighbors in every epoch. We discover that changing neighbors in every epoch usually slows down the convergence and causes the training errors vibrate a lot . Our model is a novel structure which combines both GAT and GraphSAGE, and particularly designed for link predictions. This model computes the hidden representations of each node through a shared attention mechanism across its neighbors.

A large number of experiments are implemented on five representative networks. The results show that DeepLinker not only achieve the state-of-art accuracy in link predictions but can also obtain the effective node representations for downstream tasks such as node ranking and node classifications. We find that the nodes with more attention paid by their neighbors are either the elites in a Chinese co-investment network or the 'best papers' in the APS citation network. Moreover, if the model is well trained, the low-dimensional node vectors extracted from the last layer of DeepLinker can achieve a higher node classification accuracy compared with other unsupervised learning algorithms. And the fewer the labeled nodes in classification task, the bigger the advantage our model has.

Our main contributions are summarized as follows:
\begin{itemize}
\item  We propose DeepLinker, which achieves the state-of-the-art accuracy in link prediction task.
\item We handle the memory bottleneck and mini-batch problems by fixing the neighborhood number which yields a controllable cost for per-batch computation.
\item The trained attentional coefficient matrix of DeepLinker plays a key role in revealing the latent significance of nodes. It helps in identifying the elites of a Chinese Investment Network and finding the 'best papers' for citation network.
\item DeepLinker can extract meaningful feature representations for nodes, this link prediction based node embedding achieves high accuracy in node classification task especially when the labeled nodes are sparse.
\end{itemize}

\section{GAT architectures}
\begin{figure*}
\centering
\includegraphics[scale=0.50]{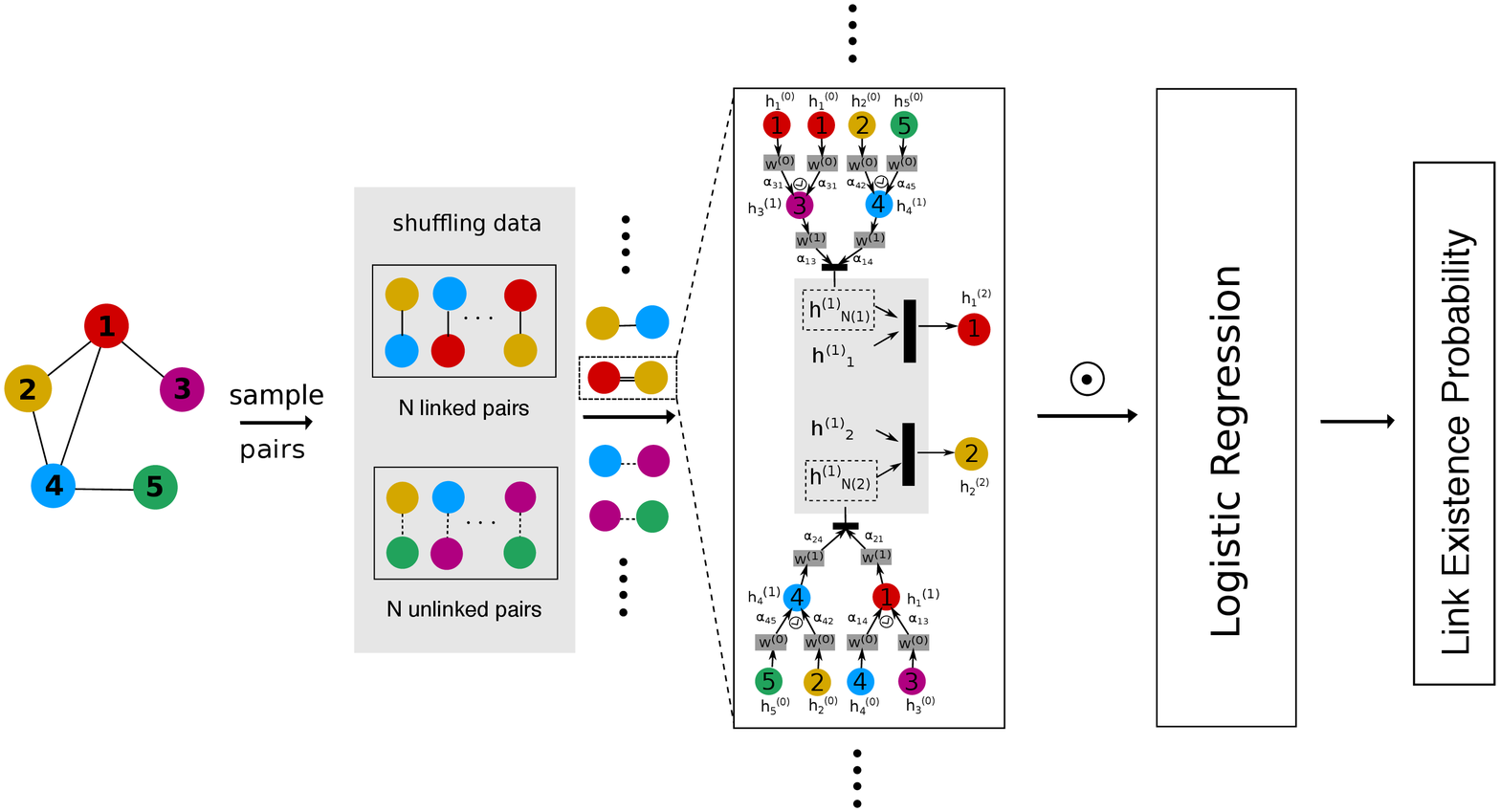}
\caption{ \label{fig:Illustration} The overall DeepLinker architecture. We take a simple five-node network as an example, any two node linkage relation will be considered. We describe the linked relations in solid lines and the unlinked ones in dashed lines. We then take the linked node 1 in red and node 2 in yellow as an training example. We first sample nodes 3 and 4 as their 1st-order neighbors, then sample nodes 1,2,5 as their 2nd-order neighbors, nodes 1,2,5 are also the 1st-order neighbors of nodes 3 and 4. After that we calculate nodes 1 and 2's vector representations based on their initial attributes as well as their neighbors' initial features via GAT architecture. We acquire edge vector via calculating the inner product of the 1 and 2's vector representations. Finally a logistic regression function is applied to compute the linkage existence probability.}
\end{figure*}

To start with, we review the architecture of GAT model which our model is mainly based on. GAT takes a set of node features as input, \textbf{h} = \{$\Vec{h}_{1}$, $\Vec{h}_{2}$, ... , $\Vec{h}_{N}$\}, in which, $\Vec{h}_{i}$ $\in$ $\mathbb{R}^F$, and $N$ is the number of nodes, $F$ is the number of input node attributes. We use $\Vec{h}_{i}' \in \mathbb{R}^{F^{'}}$ to denote GAT's outputs that also contain $F^{'}$ node features. The target of GAT is to obtain sufficient expressive power to transfer the input features into high-level output features. It first applies a learnable linear
transformation, parameterized by a weight matrix, \textbf{W} $\in$ $\mathbb{R}^{F^{'} \times F}$ to every node, it then uses a single-layer feed-forward neural
network $\Vec{a}\in$  $\mathbb{R}^{2F^{'}}$ to compute the attention coefficients between nodes. This computation process is shown in equation \ref{attention coefficient}, where $.^T$ represents matrix transposition and $\parallel$ is the concatenation operation. Node $j$ is a neighbor of node $i$, and $\alpha_{i,j}$ indicates the importance of $j$'s features to $i$ among all of $i$'s neighbors, represented as $\EuScript{N}_{i}$.
\begin{equation}
\label{attention coefficient}
 \alpha_{i,j} = \frac{\exp \left( LeakyReLU \left ( \vec{a}^{T}  \textbf {[W}\vec{h}_{i} \parallel \textbf W \vec{h}_{j}\textbf{]} \right) \right ) }{\sum_{{j \in \EuScript{N}_{i} }}\exp \left( LeakyReLU \left( \vec{a}^{T}\textbf  {[W}\vec{h}_{i} \parallel \textbf W \vec{h}_{j}\textbf{]}\right)\right)}.
\end{equation}
Once the normalized attention coefficient $\alpha_{i,j}$ is obtained, GAT aggregates nodes' features as a combination of their neighbors, followed by a potentially nonlinear sigmoid function $\sigma$, as is shown in equation \ref{nonlinear hidden layer}.
\begin{equation}
\label{nonlinear hidden layer}
\vec{h^\prime_i} = \sigma \left( {\sum_{j \in \EuScript{N}_i} \alpha_{i,j}\textbf  {W}\vec{h}_{i} } \right)
\end{equation}

Finally, GAT employs multi-head attention to stabilize the learning process of attention coefficients. \textbf{K} denotes the number of attentional heads, $\alpha_{i,j}^k$ denotes the $k$th relative attentional weights of $j's$ features to $i$. The output features of nodes' neighbors are either concatenated or averaged to form their final output features, as is shown in equation \ref{eggregator}:
\begin{equation}
\label{eggregator}
 \vec{h^\prime_i} =  \rVert_{k=1}^{\textbf{K}}  \sigma \left( {\sum_{j \in \EuScript{N}_{i}  } \alpha_{i,j}^{k}\textbf  {W}\vec{h}_{i} } \right) .
\end{equation}


\section{DeepLinker architecture}
To improve link prediction accuracy, we introduce DeepLinker which has an encoder-decoder architecture, with adjustable attention parameters. The overall architecture of DeepLink is shown in Figure \ref{fig:Illustration}. An encoder encodes nodes to a vector representations, $F^{'}$ $\in$ $\mathbb{R}^{F^{'}}$, a decoder then
genearte an edge vector by aggregating nodes' vectors. Finally, a score function is applied to evaluate the link existence probability between two nodes via the edge vectors. One of the key ideas behind DeepLinker is to learn how to aggregate nodes' features into edge vector for link prediction task.

As mentioned above, the limitations of GAT are memory bottleneck and mini-batch problem. Due to the scale-free property of most networks, once the hub nodes are sampled as the 1st-order neighbors of a given node, the 2nd-order neighbors usually quickly fill up the memory, which prevents GAT to be applied on larger networks. Besides, the existing GPU-enabled tensor manipulation frameworks are only able to parallelize on the normalized activation coefficients ($\alpha_{i,j}$) for the same sized neighborhoods, which prevents GAT from parallel computing.

\subsection{Fixed-sized Neighborhood Sampling}
Here we use the fixed-sized neighborhood sampling strategy to solve the memory bottleneck and the mini-batch problem. The undirected graph can be represented as $G=(V,E)$, with $V$ denoting the set of nodes and $E$ representing the edges in network $G$. For any two randomly selected nodes $i$ and $j$, as is illustrated in Figure \ref{fig:Illustration} we then calculate the Hadamard distance to represent the edge vector and evaluate the edge existence probability via training a logistic regression function. We sample each node's neighborhood to form a fixed-sized node mini-batch. Taking node $i$ as a sample example, we uniformly sample a fixed-sized set of its neighbors defined as $\EuScript{N}_{i}$ from the set $\left\{ i \in V: (i,j) \in  E  \right\}$ instead of full neighborhood set as in GAT. Different from sampling neighborhood during each training iteration in GraphSAGE, we sample only once and fix the neighborhoods during the whole training process. The sampling strategy is illustrated in Figure \ref{fig:Illustration}.
We compute the Hadamard product between two output vectors to generate an edge vector, as is shown in Equation \ref{Hadamard product}, note that $e_{ij}$ is a $d$-dimensional vector.
\begin{equation}
\label{Hadamard product}
e(i,j) = (\vec{h^\prime_i} \odot \vec{h^\prime_j})
\end{equation}
We then assume that the probability between node $i$ and $j$ is given by Equation \ref{link probability} where $\theta$ is a $d$-dimensional parameter vector, and $e_{ij}^{T} \theta$ is the dot product between the vectors $e_{ij}$ and $\theta$.
\begin{equation}
\label{link probability}
p_{ij}(e_{ij};\theta) =  \frac{1}{1+exp(e_{ij}^{T} \theta)}
  \end{equation}
\subsection{Training DeepLinker}
The whole framework is trained by minimizing the following objective function:
\begin{equation}
\label{loss function}
    \mathcal{L} =  -\frac{1}{\mid\varepsilon\cup \varepsilon^{-}\mid} \sum_{(i,j)\in \varepsilon\cup \varepsilon^{-}} y_{i,j}\log p(i,j) +(1-y_{i,j})\log(1-p(i,j)),
\end{equation}
where, $y_{i,j}$ is the label information for linkage between $i$ and $j$, with $0$ for non-existence and $1$ for existence. $\varepsilon$ is the training set of edges. We follow the convention to randomly divide the full set of edges $E$ into three parts, $\varepsilon$ for training, $\phi$ for validating and $\tau$ for testing. We train the model to predict not only the existence but also the non-existence links. Here, $\varepsilon^{-}$ is the set of negative samples, in which each element is a node pair $(i,z)$, and both $i$ and $z$ are drawn from the nodes involved in $\varepsilon$. Additionally, there is no edge between $i$ and $z$ in the set of $\varepsilon^{-}$. In our experiment, we sample the same number of negative instances as for positive links.


\section{Experiments}
In order to evaluate the performance of DeepLinker and explore its potential applications, we conduct link prediction, node ranking, as well as node classification tasks over several networks ranging from citation network to venture capital co-invest network. All experiments are done under the PyTorch machine learning environment with a CUDA backend.
\subsection{Experiments setup}
We utilize the following five networks in our experiments:
\begin{itemize}
\item Cora network contains 2,708 nodes 5,429 edges, and 7 classes. Each node has 1,433 attributes corresponding to elements of a bag-of-words representation of a document.
\item Citeseer network contains 3,327 nodes 4,732 edges, and 6 classes. Each node has 3,703 attributes extracted from paper contents.
\item Pubmed is a citation network, which contains 19,717 nodes and 44,338 edges, and 3 classes. The size of attributes of each node is 500.
\item VC network is a venture capital co-invest network with nodes representing venture capital companies and edges corresponding to co-invest events. It contains 1,436 nodes and 2,265 edges. Here we use the adjacency matrix as the one-hot input node attribute in the training process. On this network, 42 nodes are identified manually as VC which play a vital role in venture capital events. These nodes are regarded as the ground truth in node ranking task. The VC investment network is built on the SiMuTon \cite{zero2ipo} database.
\item APS graph has 1,012 nodes and 3,336 edges. The adjacency matrix is used as the the one-hot input node attribute in the training process. Follow a paper published on Science \cite{wang2013quantifying} here we quantify papers' impact and importance by counting the number of citations over 10 years (c\_10) after their publication, c\_10 is used as ground truth metric for measuring nodes importance.
\end{itemize}
We compare DeepLinker with the following baseline algorithms:
\begin{enumerate}
\item RA\cite{zhou2009predicting} is a traditional link prediction method, and the similarity between two nodes is measured by computing neighbors' weights which are negatively proportional to its degree.
\item LINE\cite{tang2015line} minimizes a loss function to learn embedding while preserving the first and the second-order neighbors proximity among vertices in the graph.
\item Node2vec\cite{grover2016node2vec} adopts a biased random walk strategy and applies Skip-Gram to learn vertex embedding. This embedding algorithm is widely used in recent years.
\item GraphSAGE\cite{hamilton2017inductive} learns node embedding through a general inductive framework consisting with several feature aggregators. It usually adopts supervised node classification task as the evaluation benchmark with the assumption that better embedding algorithm leads to higher node classification accuracy.
\item GAT\cite{velickovic2017graph} is the main architecture that our model based on. GAT compute representation of each node by combining its neighborhoods vectors in an adaptive way with adjustable attention weights for different neighborhoods.
\end{enumerate}

In our experiments, we keep node2vec and LINE's parameters as they are in the original papers. We set the same parameter for GraphSAGE, GAT and DeepLinker, that include the type and sequence of layers, choice of activation functions, placement of dropout, and setting of hyper-parameters.

DeepLinker algorithm consists of two layers, the first layer is made up of 8 ($K1 = 8$) attentional heads over all networks. Here we set the hidden size to 32 for Cora and VC, 16 for Citeseer and APS, and 64 for Pubmed network. The main purpose of the first layer is to compute the hidden features of the 1st-order neighbors. We then add the non-linearity by feeding the hidden features to an exponential linear unit (ELU), as shown in equation \ref{nonlinear hidden layer}. The aggregated features from each head are concatenated in this layer.

The main purpose of the second layer is to compute the edge features that used for evaluating link probability. Here we use a single attention (K2 = 1) for Cora, Citeseer, VC and APS graphs. We discover that the Pubmed graph requires a much larger attention head number, thus we set (K2 = 8) for Pubmed graph. The aggregated features from each head are averaged in this layer. The output of the second layer is the final feature representations for nodes. We then compute the Hadamard distance between two node features to represent the edge vector, as is shown in equation \ref{Hadamard product}. Once edge vectors are obtained, an active function sigmoid $\sigma$ is applied to evaluate the edge existing probability between nodes.

We initialize the parameter of DeepLinker with Glorot initialization \cite{glorot2010understanding} and train to minimize the
binary-cross-entropy\ref{loss function}, for the training set we use the Adam SGD optimizer\cite{kingma2014adam} with an initial learning rate of $5e-4$. We also apply an early stop strategy over link prediction accuracy for the validation set, with the patience sets to 100 epochs.

We solve the memory bottleneck by sampling a fixed neighbors size (20) for both 1st and 2nd-order neighbors selection for DeepLinker. The sampling strategy is illustrated in Figure \ref{fig:Illustration}. We set the batch size to be 32 for Cora, Citeseer, and Pubmed, 16 for APS network, and 8 for VC network. The sampling size in GraphSAGE algorithm is also 20.

\subsection{Link prediction}
In this part, we evaluate the link prediction accuracy of DeepLinker algorithm and compare DeepLinker with other link prediction algorithms mentioned above. The goal of link prediction task is to predict whether there exists an edge between two given vertices. To start with, we randomly hide 10\% of the edges in the original graph to form 'positive' samples in the test set. The test set also has equal number of randomly selected disconnected 'links' that servers as 'negative' samples. We then use the remaining 90\% connected links and randomly selected disconnected ones to form the training set. After that, we uniformly sample first and second order neighbors for each node. Finally we feed the sampled nodes into DeepLinker and the output of DeepLinker is the edge existence probability between two vertices.

Two standard metrics, Accuracy and AUC (area under the curve) are used to quantify the accuracy of link prediction. As shown in Table \ref{tab:accuracy}, DeepLinker outperforms all the baseline methods in link prediction accuracy across all graphs. In Table \ref{tab:accuracy} we didn't report the link prediction accuracy of GAT algorithm, because Citeseer and Pumbed datasets caused memory error in our machine.

\begin{table}
\centering

\caption{Link prediction accuracy and AUC of different algorithms over several networks. }
\label{tab:accuracy}
\begin{tabular}{cccccc}
\hline
Accuracy/AUC             & Cora       & Citeseer    & Pubmed     & VC network  & APS network\\ \hline
GAT                   & 0.79/0.88  &  NA          &  NA         & 0.77/0.83   & 0.80/0.89\\
DeepLinker (attention) & 0.87/0.93  & 0.85/0.91   & 0.57/0.63  & 0.80/0.90   &0.84/0.94\\
DeepLinker (all ones)   & \textbf{0.88/0.93}  & \textbf{0.86/0.91}   & \textbf{0.90/0.97}  & \textbf{0.82/0.90} &\textbf{0.85/0.95}  \\
GraphSAGE-mean                  & 0.83/0.89 & 0.84/0.90   & 0.88/0.96 & 0.81/0.87 & 0.82/0.88 \\
node2vec              & 0.82/0.92 & 0.85/0.89  & 0.81/0.94 & 0.77 /0.87 & 0.83/0.89 \\
LINE                  & 0.69/0.76  & 0.67 / 0.73 & 0.66/0.72 & 0.78/0.84 & 0.68/0.74 \\
RA                    & 0.41/0.75  & 0.32/0.73   & 0.31/0.69  & 0.33/0.76   &0.35/0.78 \\ \hline
\end{tabular}
\end{table}

Here, we propose two implements of DeepLinker: DeepLinker (attention) and DeepLinker (all ones). Both implements apply neighborhood sampling and mini-batch strategies during the training process, while DeepLinker (attention) trains the attention coefficient, which indicates how important the neighbors' features to the present node, as illustrated in equation \ref{attention coefficient} and DeepLinker (all ones) sets the attention to 1 among all neighbors. As is shown in Table \ref{tab:accuracy}, DeepLinkerDeepLinker (attention) has a much higher prediction accuracy compared with other link prediction algorithms, while DeepLinker(all ones) achieves the best performance among all datasets. Actually in larger networks such as Pubmed, no matter how we adjust the learning rate, the linear transformation matrix size and the mini-batch size for the attention based DeepLinker, the training error doesn't go down. At the initial stage of the training process, the neighbors' attributes have a stronger influence on the representation ability of the current node, however the attention coefficient value in DeepLinker (attention) architecture can only apply a limited attention coefficient ranging from 0 to 1, which weakens the neighbors' feature influence on the current node. Attention coefficient tends to be less than 0.1 especially for a hub node with tremendous neighbors, in other words, the attention mechanism limits the expression power on neighbors features' influence for the current node.

DeepLinker (all ones) sets all the attention coefficients $\alpha_{ij}$ to 1, which means all neighbors' features have an equal contribution to the present node. Although DeepLinker (all ones) is a much simpler architecture with less trainable parameters, to our surprise, its performance is even better than DeepLinker (attention) over all networks, and the accuracy difference becomes much significant on large networks, such as Pubmed, as is shown in Table\ref{tab:accuracy}. When training the Pubmed graph with DeepLinker (all ones), the training error converges within 10 epochs. Compared with DeepLinker (attention), DeepLinker (all ones) is a more suitable model for link predictions on large networks.

Table\ref{tab:accuracy} shows that RA and LINE algorithms may can not capture the essential pattern of graph structure, since the predictive accuracies are low in both algorithms. Node2vec performs better than LINE and RA since the Skip-Gram model is better at extracting neighborhood information from graph structure. The original GraphSAGE-mean and GAT model are used for node classification task only, here by computing the Hadmard distance between two nodes and adding a logistic regression layer we make those models suitable for link prediction task. Since there are no sampling and mini-batch training strategies in the original GAT model, thus, once the network becomes large, the original GAT algorithm suffers from memory bottlenecks. That is why we do not report the link prediction accuracy on the Citeseer and Pubmed networks of the GAT model.

In order to test the robustness of the model, we randomly break 20\% edges among the existing links. Table \ref{tab:robustness} shows that both DeepLinker (attention) and DeepLinker (all ones) are much robuster than other algorithms, and DeepLinker (all ones) achieves the state-of-the-art link prediction accuracy.
\begin{table}
\centering
\caption{\label{tab:robustness} Link prediction robustness test on Cora graph.}
\begin{tabular}{cc}
\hline
Break Portion 20\% & Accuracy/AUC  \\ \hline
     DeepLinker (attention)      &     0.84/0.90      \\
     DeepLinker (all ones)       &         \textbf{0.85/0.91}  \\
   GraphSAGE-mean         &       0.81/0.90    \\
   node2vec          &      0.80/0.89     \\
      LINE     &        0.52/0.53   \\
     RA      &          0.33/0.73 \\ \hline
\end{tabular}
\end{table}

\subsection{Attention coefficients for node centrality measuring}
DeepLinker (all one) model is easier to train and achieves better performance in link prediction task compared with DeepLinker (attention) model. Then what's the meaning to train the attention coefficients of networks? In this part we show that the learned attention coefficients from DeepLinker (attention) help extract the hidden relationship between connected nodes for a given graph. Based on network structure DeepLinker (attention) can automatically learn a directed link weighted for any connected nodes, thus turn the undirected and unweighted networks to a directed and weighted networks. Besides, the learned attention coefficients also help in node centrality measuring and ranking. We show the attention coefficients' power in ranking and finding the most vital nodes with the Chinese Venture Capital (VC) and the APS citation networks. The detailed network information is described in the Experiments setup section.

One of the most important questions in venture capital analysis field is to identify the leading investors (leading VC) among a large amount of invest activities. Syndication in the Chinese venture capital market is typically led by main leading VCs, who always finding good investment opportunities, setting up investment plans, and organizing the partners. These leaders play a major role during the investment activities, therefore, identifying them has practical significance. To find the ground truth for discovering and identifying the leading VCs in venture capital industry, we use the Delphi method to interview four experts in this field to get a name list of leaders among the VC firms \cite{gu2019exploring}. Based on this questionnaire survey, we identify 42 elites (leading VCs) in this network.

The APS graph is a sub-graph extracted from APS (American Physical Society journals) website with nodes representing papers and links representing citations. Measuring the centrality of papers helps scientists find the significant and high quality discoveries among thousands of publications. In this paper, we follow a paper published in Science journal\cite{wang2013quantifying} to evaluate papers' importance by counting the number of citations within the first 10 years (c\_10) after papers' publications.

Intuitively, the more attention a node attracts, the more influential it is. We measure node's influence by accumulating neighborhoods' attention towards it across all heads in the second layer of DeepLinker(attention). We name this attention coefficient based node rank method as Attention Rank, as is shown in equation \ref{ranking equation}.
Attention Rank is a byproduct of DeepLinker (attention). We first extract the normalized attention coefficients $\alpha_{j,i}^k$ from a pre-trained DeepLinker (attention) model of the $k^{th}$ attentional head in the second layer, and then sum them up for all neighbors across all heads.
\begin{equation}
\label{ranking equation}
 \bar{c}_{i} =  {\sum_{k=1}^{K_2}\sum_{j \in \EuScript{N}_i} \alpha_{j,i}^k }
\end{equation}
By calculating VC's total amount of attention based on equation \ref{ranking equation}, we find that the elites (leading VCs) always attract a larger amount of attention compared with the followers. Following the evaluation methods for node centrality measures in complex networks \cite{aral2012identifying}, we sort VC nodes according to the total amount of attention in a decreasing order, and find that we can hit 30 elites out of the top 42 elites in the ground truth set. Table \ref{vc rank table} shows the top 16 VCs with the most attention, all of the VCs are elites in the ground truth set. Besides, we find the top 16 VCs has a larger overlaps with a later released VC ranking website that discusses about the 'best' venture capital in China\cite{bestVc}.

\begin{table}
\centering
\caption{\label{vc rank table} The top 16 VC firms that attract the most attention.}
\begin{tabular}{cccccc}
\hline
Rank & VC name                       & is\_elite & Rank & VC name             & is\_elite \\ \hline
1    & MORE/Shenzhen Capital Group    & Yes   & 9    & JAFCO ASIA           & Yes   \\
2    & IDG Capital                    & Yes   & 10   & FOETURE Capital      & Yes   \\
3    & Sequoia                        & Yes   & 11   & GGV Capital          & Yes   \\
4    & Legend Captital                & Yes   & 12   & Walden International & Yes   \\
5    & Goldman Sachs                  & Yes   & 13   & SBCVC                & Yes   \\
6    & Intel Capital                  & Yes   & 14   & DFJ Venture Capital  & Yes   \\
7    & Northern Light Venture Capital & Yes   & 15   & Qiming               & Yes   \\
8    & DT Capital                     & Yes   & 16   & Cowin                & Yes   \\ \hline
\end{tabular}
\end{table}

To accurately compare the ranking result with other centrality measuring algorithms, we follow the method in Webspam competition \cite{heidemann2010identifying} and use the Accuracy which defines the ratio of the hits on the ground truth as a metric to evaluate the performance of different node ranking methods. The ranking performances are listed in Table \ref{tab:ranking}.

In order to evaluate papers' importance for APS citaion network, we first extract the pre-trained attention coefficient matrix from DeepLinker and rank papers' total attention in a decreasing order. We then follow the experiments of \cite{wang2013ranking} to use Spearman's rank correlation coefficient as an evaluation metric since the ground truths (c\_10) are real values instead of binary ones.

In this part we choose several unsupervised graph ranking algorithms to compare with. The first algorithm is PageRank \cite{page1999pagerank} which is a basic solution for ranking nodes. The second is Closeness Centrality and the third is Betweenness Centrality \cite{freeman1978centrality}. The Closeness Centrality believes that the most important nodes should have shorter path lengths to other nodes, while the Betweenness Centrality assumes that the most important nodes should be involved in more shortest paths. We also compared Attention Rank with the SPRank \cite{zhou2016ranking}, which can efficiently and accurately identify high quality papers (e.g. Nobel prize winning papers) and can significantly outperform PageRank in predicting the future citation growth of papers. Table \ref{tab:ranking} shows the compare result of DeepLinker (attention) with the ranking methods mentioned above with the default parameter settings. We can find that Attention Rank significantly outperforms other ranking methods without any adjustable parameter and human knowledge.

\begin{table}
\caption{\label{tab:ranking}Ranking performance comparison under unsupervised ranking methods.}
\begin{tabular}{ccccccc}
\hline
Dataset & Evaluation & PageRank & Closeness & Betweenness & SPRank & Attention Rank \\ \hline
VC      & Accuracy   & 0.65     & 0.60      & 0.58        &0.64       & \textbf{0.72}   \\
APS     & Rank Coor. & 0.32     & 0.08      & 0.03        &0.38       &\textbf{0.42}    \\ \hline
\end{tabular}
\end{table}

\subsection{Feature learning for node classification}
\begin{figure*}
\begin{minipage}[t]{0.33\linewidth}
\centering
\includegraphics[width=2.7in,height=2.2in]{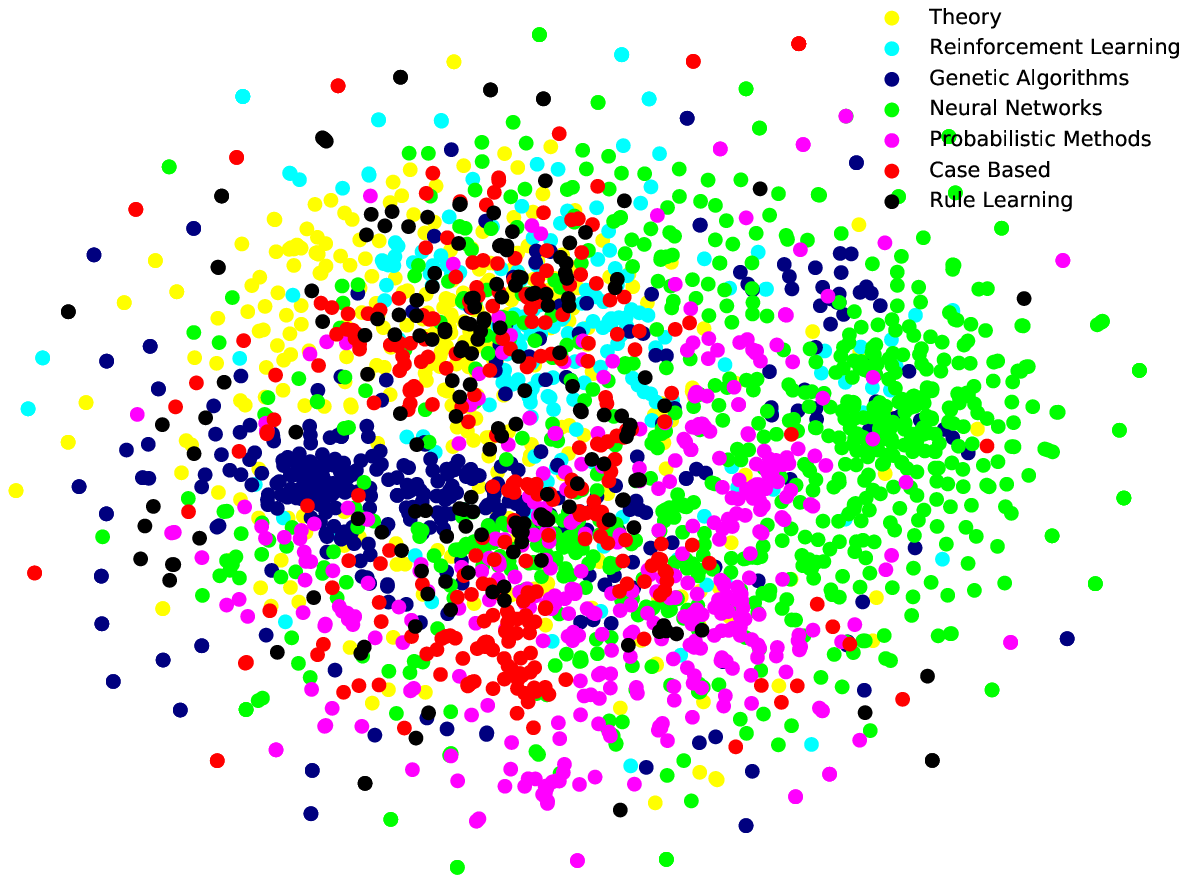}
\textbf{Raw feature visualization}
\end{minipage}%
\begin{minipage}[t]{0.33\linewidth}
\centering
\includegraphics[width=2.7in,height=2.2in]{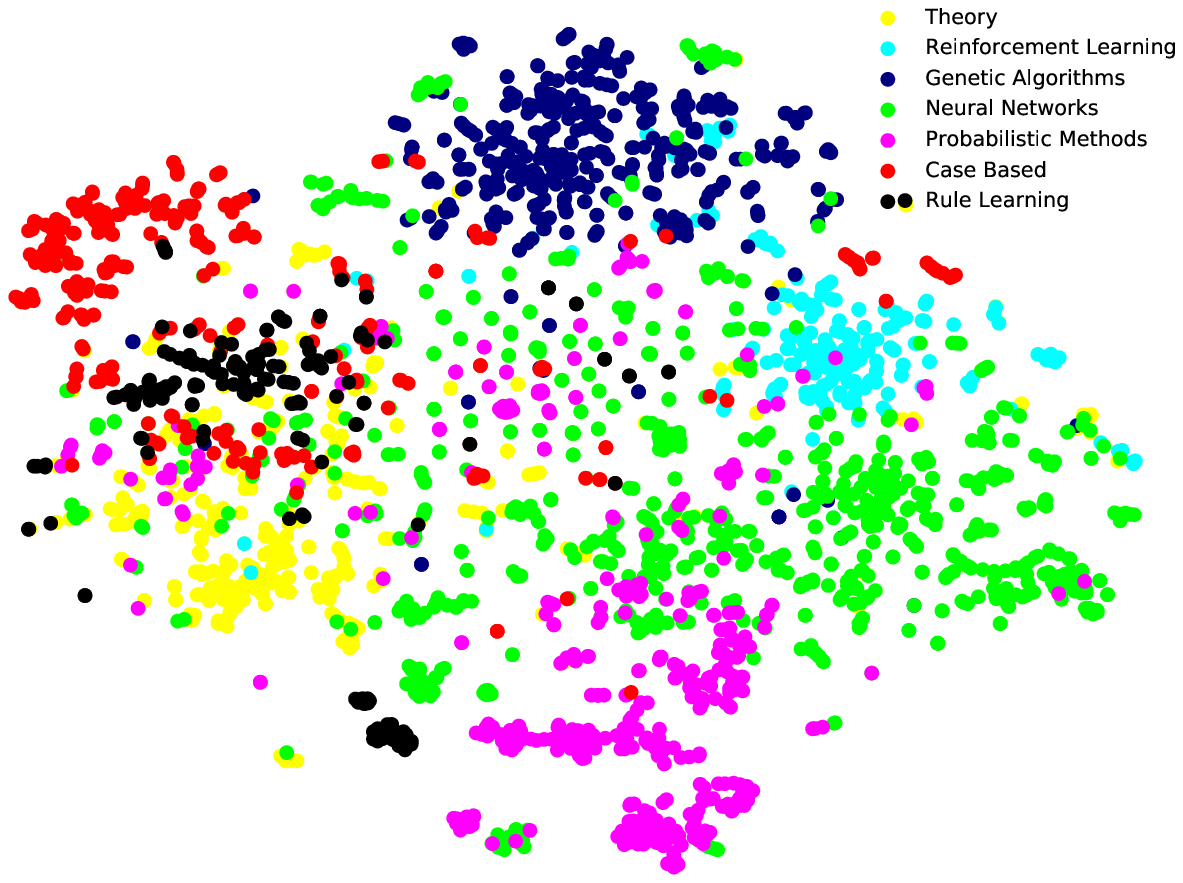}
\textbf{node2vec visualization}
\end{minipage}
\begin{minipage}[t]{0.33\linewidth}
\centering
\includegraphics[width=2.7in,height=2.2in]{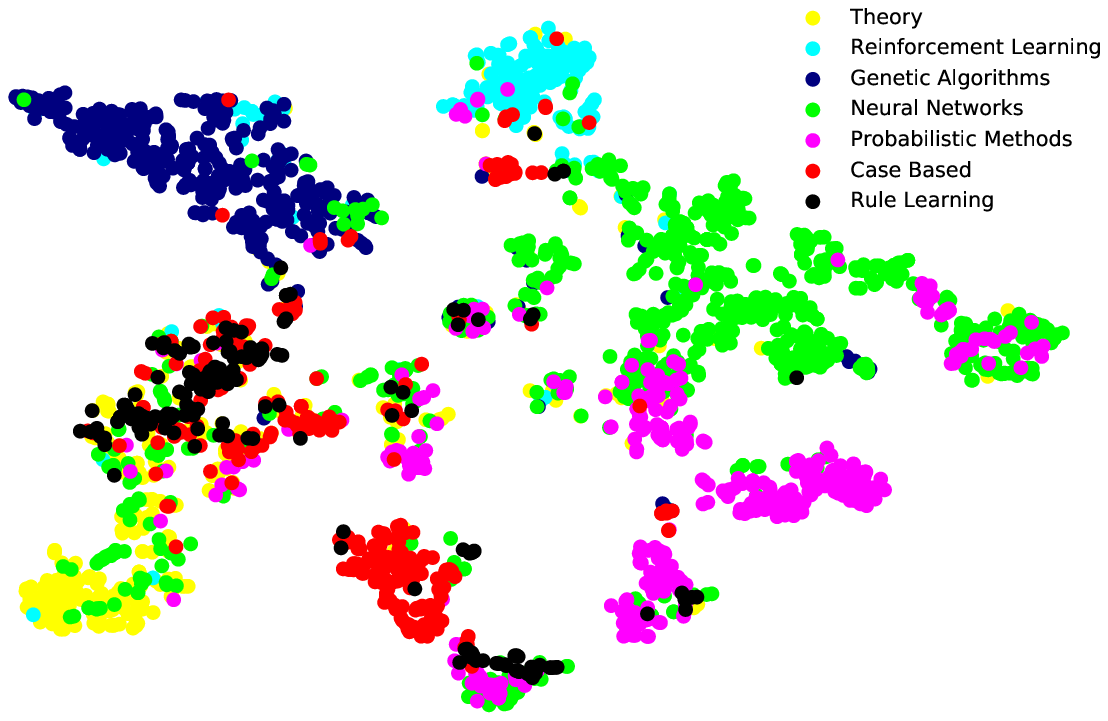}
\textbf{DeepLinker visualization}
\end{minipage}%
\caption{ \label{fig:subplot} t-SNE visualization of Cora graph from the raw features (left), node2vec representations with the default parameter setting (middle), and the DeepLinker representations, node features for DeepLinker are extracted from the second layer of a pre-trained model (right). The clusters of the DeepLinker's representations are clearly defined with a Silhouette score equals to 0.38 compared with 0.09 in node2vec and 0.00 in raw features. The NMI  value of DeepLinker is 0.44 compared with 0.41 in node2vec vectors and 0.13 in raw features.}
\end{figure*}

DeepLinker can not only be applied in predicting missing links, measuring node centrality but also can provide meaningful node representations. In Figure \ref{fig:subplot} we visualize the raw input attributes, the second layer output of the pre-trained GraphSAGE-mean and the pre-trained DeepLinker (all ones) with t-SNE \cite{maaten2008visualizing} visualization method. In this figure each point represents a node in the Cora graph with its node color denoting classification label. From the DeepLinker (all ones) visualization, we can tell in general, nodes belonging to the same class usually block together. The NMI (Normalized Mutual Information) and the Silhouette score of DeepLinker representation are much higher than GraphSAGE-mean vectors and raw attributes. For example, the Reinforcement Learning and the Genetic Algorithms are quite independent communities with the highest sub-network density 0.017 and 0.009 compared with the whole network's density (0.001). This indicates that link prediction based representation has incorporated node similarity information. We also discover that even DeepLinker tends to separate nodes from different classes, there are still some overlap areas consisting of nodes from different classes. This phenomenon may support the popularity versus similarity theory~\cite{papadopoulos2012popularity}, which claims that network links are trade-offs between node similarity and other network properties such as node popularity.

Compared with the network representations learned from supervised node classification task such as GAT, GCNs, and GraphSAGE, the representations learned from links such as DeepLinker and node2vec contains richer structural information. The node representations learned from supervised node classification can only decode part of the hidden information. For example in citation networks, papers belonging to the same subject in general will have the same labels. Network embedding based on node labels would decode mainly the similarity between nodes. However, the evolution of citation networks rely not only on papers' subjects and labels, other factors such as authors' fame and their popularity also play vital roles in network formation \cite{hunter2011dynamic}. Besides, the supervised link information is easier to acquire compared with the supervised label information. Actually, there are few networks that have node label information.

In order to evaluate the effectiveness of the proposed DeepLinker representation, we follow the commonly adopted setting to compare different representation algorithms of node classification task on Cora and Citeseer networks. In node classification task, each node has a targeted label and our goal is to build a simple predictive learning logistic model based on the training set and to predict the correct labels in the test set. In particular, after extracting nodes representations of a given graph, we randomly select some nodes to form the training data, we then apply the training nodes representations and their labels to train a node classify model and use this model to predict the label of the remaining nodes. We repeat the classification experiments for 10 times and report the average Micro-F1 value.

\begin{figure*}
\begin{minipage}[t]{0.45\linewidth}
\centering
\includegraphics[width=3.5in,height=2.7in]{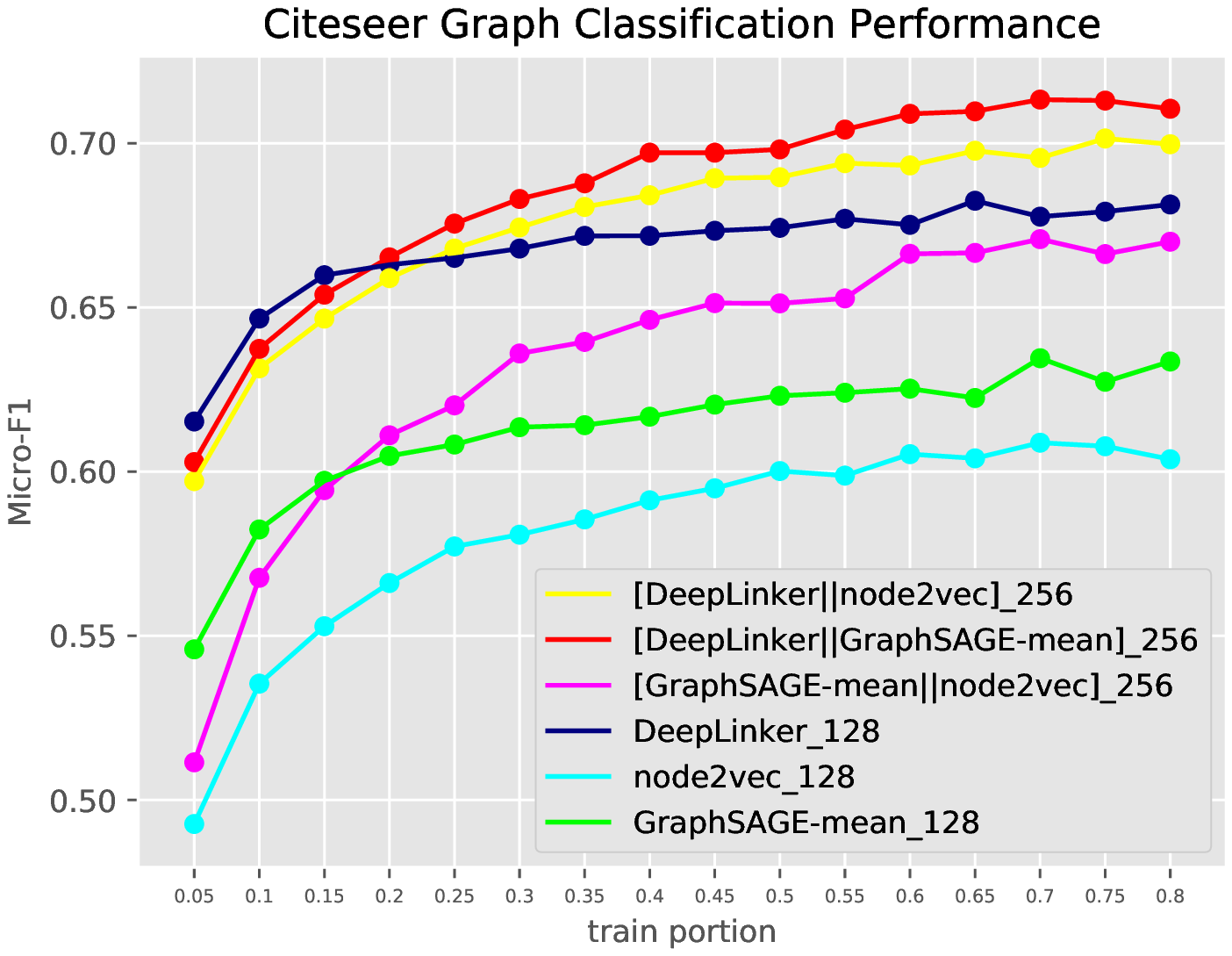}
\end{minipage}%
\begin{minipage}[t]{0.45\linewidth}
\centering
\includegraphics[width=3.5in,height=2.7in]{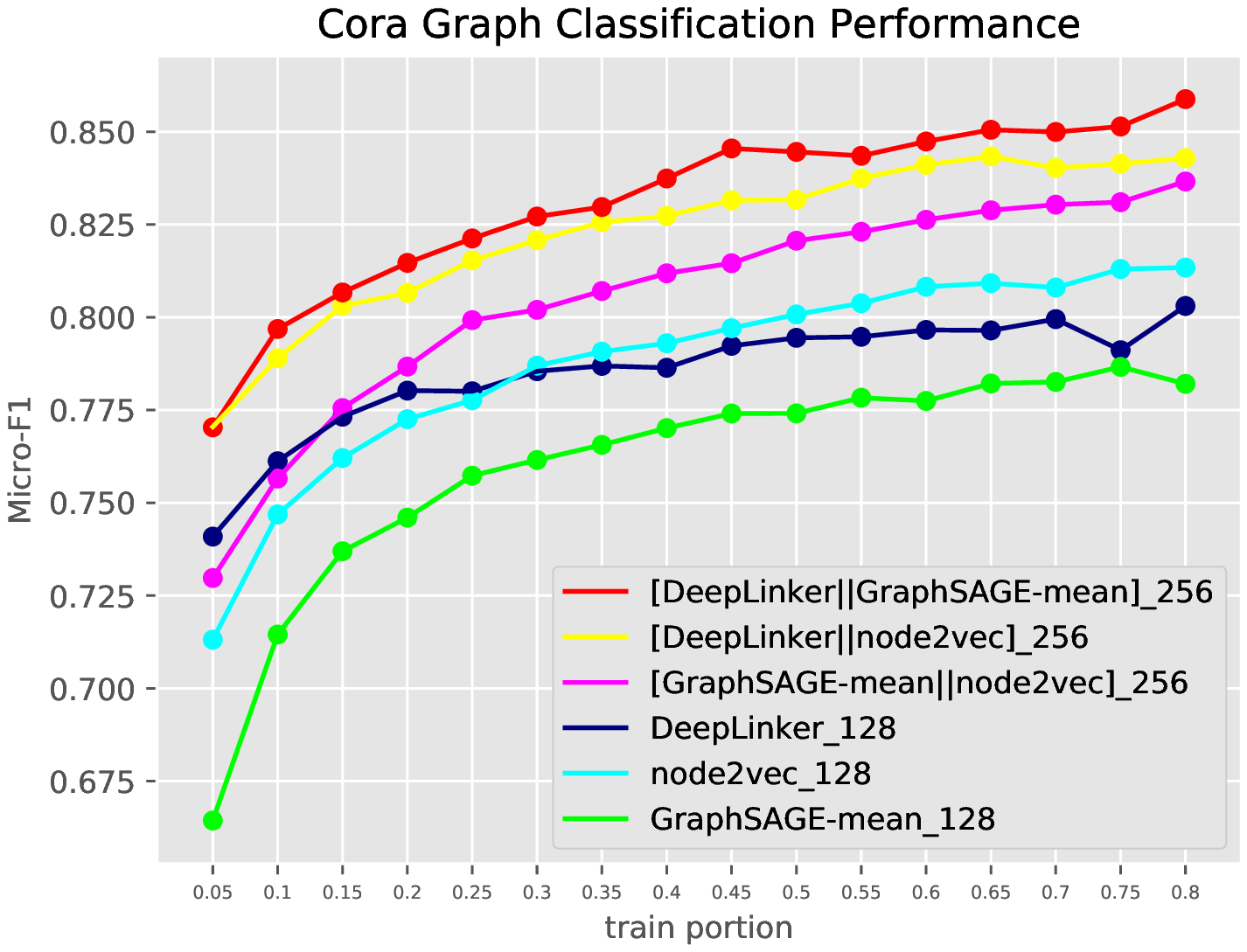}
\end{minipage}%

\caption{\label{fig:nodeclassification} Classification accuracy comparison under different representation learning methods.}
\end{figure*}

We compare the node vectors learned from link prediction task of DeepLinker (all ones) with the unsupervised GraphSAGE-mean and the widely used $node2vec$. In order to control changing variables, we fix the embedding dimension to 128 for all algorithms, and name them as DeepLinker\_128, GraphSAGE-mean\_128 and node2vec\_128. The first sub-graph in Figure \ref{fig:nodeclassification} shows that in Citeseer graph, DeepLinker outperforms $node2vec$ and GraphSAGE-mean especially when the labeled training set is small. In the second sub-graph, when the training portion is less than 10\%. DeepLinker also performs better than $node2vec$ and GraphSAGE-mean. Here, we believe achieves a better performance when the training set is small is an important factor in node classification task, because in real-world networks there are only a few labeled graphs and manually labelling a large amount of nodes not only cost time and efforts but also introduce biases.

Moreover, in order to improve node classification accuracy, we increase the embedding dimension by concatenating the vectors learned from different unsupervised algorithms. As is shown in \ref{fig:nodeclassification}, we find the concatenation of DeepLinker and GraphSAGE-mean achieves the highest classification accuracy in Cora and Citeseer compared with other combination, the combination of DeepLinker and $node2vec$ also has a better classification accuracy compared with the joint of GraphSAGE-mean and $node2vec$. Actually the concatenating of GraphSAGE-mean and node2vec performs even worse than DeepLinker in 128 dimension in the Citeseer network.

\section{Conclusion and discussion}
In this paper, we propose a mini-batched link prediction model, DeepLinker which based on the graph convolution architecture and sampling strategy. DeepLinker can extract meaningful vertex representation and achieve the state-of-the-art link prediction accuracy. The byproducts of DeepLinker, attention coefficients show the potential in node centrality measuring and  the node representations extracted from the last layer show advantage in node classification tasks, especially when the labeled training data is small. DeepLinker outperforms other unsupervised node representation learning methods in node classification and node visualization tasks. This may alleviate the dependency on large labeled data.

Despite the process of adjusting the hyper-parameter requires many efforts, we still believe that network representation based on link prediction can lead to both quantitative and qualitative leap in graph processing. Although DeepLinker has achieved a quite high accuracy of link prediction, we still can't figure out the mechanism that leads to such a good performance. Our future work will mainly focus on the hidden theory behind DeepLinker.



\end{document}